\documentclass[10pt,letter,journal]{article}
\usepackage{amsmath,amssymb}
\usepackage{latexsym}
\usepackage[dvips]{pstricks} 
\usepackage[dvips]{graphicx}
\usepackage{pst-node}

\newcommand{\N}{\mathbb{N}}
\newcommand{\R}{\mathbb{R}}
\newcommand{\Z}{\mathbb{Z}}

\renewcommand{\j}{\mathbf{j}}
\newcommand{\e}{\mathbf{e}}

\newcommand{\n}{\mathbf{n}}
\newcommand{\p}{\mathbf{p}}
\newcommand{\q}{\mathbf{q}}

\newcommand{\1}{\mathbf{1}}

\newcommand{\rC}{\rm{C}}

\newcommand{\lan}{\langle}
\newcommand{\ran}{\rangle}
\newcommand{\an}[1]{\lan#1\ran}
\newcommand{\hs}{\hspace*{\parindent}}
\newcommand{\proof}{\hs \textbf{Proof.\ }}

\newcommand{\trans}{^\top}
\newcommand{\qed}{\hspace*{\fill} $\Box$\\}

\newcommand{\Gam}{\mathbf{\Gamma}}

\newcommand{\perio}{\mathrm{per}}

\newtheorem{theo}{\bfseries \hs Theorem}[section]

\numberwithin{equation}{section}

\begin{document}

\title{The $1$-vertex transfer matrix and accurate estimation of
channel capacity}

\author{
S. Friedland\thanks{Department of Mathematics, Statistics and Computer Science,
University of Illinois at Chicago, Chicago, Illinois
60607-7045, USA,
$<$friedlan@uic.edu$>$.},
P.H. Lundow\thanks{Department of Theoretical Physics, AlbaNova University
Center, KTH, SE-106 91 Stockholm, Sweden, $<$phl@kth.se$>$.} and
K. Markstr\"om\thanks{Department of
Mathematics and Mathematical Statistics, Ume\aa University, SE-901
87 Ume\aa, Sweden, $<$Klas.Markstrom@math.umu.se$>$.}}

\date{May 4, 2009}

\markboth{IEEE Transactions on INFORMATION THEORY }%
{The $1$-vertex transfer matrix and accurate estimation of
channel capacity}

\maketitle


\begin{abstract}
    The notion of a $1$-vertex transfer matrix for multi-dimensional
    codes is introduced.  It is shown that the capacity of such codes, or 
    the topological entropy, can be
    expressed as the limit of the logarithm of spectral radii of
    $1$-vertex transfer matrices.  Storage and computations using the
    $1$-vertex transfer matrix are much smaller than storage and
    computations needed for the standard transfer matrix.
    
    The method is applied  to estimate the first $15$ digits of the
    entropy of the 2-dimensional (0,1) run length limited channel.  
    
    In order to compare the computational cost of the new method with the
    standard transfer matrix and have rigorous bounds to compare the
    estimates with a large scale computation of
    eigenvalues for the (0,1) run length limited channel in 2 and 3
    dimensions have been carried out. This in turn leads to improvements on 
    the best previous lower and upper bounds for that channel.
    \\[\baselineskip] 2000 Mathematics Subject 2000 Mathematics
    Subject Classification: 05A16, 28D20, 37M25, 82B20
    \\[\baselineskip] Keywords and phrases: Channel capacity,
    transfer matrices, multi-dimensional codes, optical storage.
\end{abstract}

\section{Introduction}
In classical information theory the emphasis was on one-dimensional
channels. This is the natural channel type when one is considering
sequential information transfer and information storage on
effectively one-dimensional media, such as a magnetic tape. However, in 
modern information technology higher dimensional channels are
becoming more and more important. The simplest example is of course
information storage on a surface, as on a CD or DVD. Future devices
seem likely to move on to three-dimensional storage media as well.

Just as for the classical one-dimensional channels one still wishes to
construct codes to protect data from errors and to find the channel
capacity given by the code.  In higher dimensions computing the exact
capacity becomes much more difficult and even for very simple codes we
do not know the exact value.  The typical example is the run-length
limited channel.  Consider a two-dimensional square grid where each
point can be assigned a value of either 0 or 1 with the restriction
that two points which differ by one in exactly one coordinate cannot
both be 1.  This channel has been studied in both information theory
\cite{NaZ}, combinatorics \cite{CW} and physics \cite{BET}.  In
statistical physics this has been studied in terms of lattices gases,
primarily through non-rigorous methods and simulation.  However a
sequence of authors \cite{CW,KZ,FP,NaZ,NaZ2} have developed methods for establishing
rigorous upper and lower bounds by using transfer matrices.  The
drawback with these methods is that the cost of achieving a given
precision is exponential in terms of the precision.  It is possible to
use symmetries \cite{LM} to speed up the computation but typically the
cost remains exponential, with improved constants.

Hence it would be desirable to be able to establish good estimates for
the capacity of a code before deciding whether to invest the time
needed to bound its capacity rigorously.  In this paper we present a
method for obtaining such estimates.  The method uses a so called
1-vertex transfer matrix which adds one vertex at a time to the graph,
rather than whole rows as in the standard method.  We prove that under
certain conditions on the code the largest eigenvalue of this matrix
converges to the channel capacity.  The drawback is that we do not get
easily computable bounds for the capacity, but as a practical example
will demonstrate one can get estimates which agrees with the rigorous
bounds at a fraction of the computational cost.  Hence this method can
be used as a tool for quickly testing many different candidate codes in
order to  identify promising coding schemes, for which one can later 
establish rigorous bounds as well.

In order to help evaluate the accuracy our new method we have also
improved the known bounds for the capacity of the two and three-dimensional
run-length limited channels, or hard core-lattice gases as they are
known in the physics literature.

\subsection{Mathematical preliminaries}
Many channels and models from statistical physics can be described in
terms of restricted colourings of an underlying graph, often a lattice
such as $\Z^{d}$, or more generally in terms of weighted graph
homomorphisms \cite{BCL, FLS,LM}.

We now explain the main ideas and results of our paper for the two-dimensional 
grid $\Z^2$, where the colouring conditions are
\emph{symmetric} and \emph{isotropic}.  Let $\Delta=(V,E),
V=\{1,\ldots,k\}$ be an undirected graph on $k$ vertices, where vertex
$i$ is identified with the colour $i$.  Two neighbouring points $\p=(
p_1,p_2)^{\trans}, \q=(q_1,q_2)\trans\in\Z^2$, i.e.
$|p_1-q_1|+|p_2-q_2|=1$, can be coloured in colours $i,j\in V$ if and
only if $(i,j)\in E$.  We assume that $\Delta$ has no isolated
vertices.

The standard transfer matrix is defined as follows.  For a positive
integer $n\in \N$ denote by $\an{n}=\{1,\ldots,n\}$.  Then
$\phi:\an{n}\to V$ is viewed as a colouring of $\an{n}$ in $k$ colours.
View $\an{n}$ as $n$ points on the real line, where $i,j\in\an{n}$ are
neighbours if and only if $|i-j|=1$.  Then
$\phi\in\rC_{\Delta}(\an{n})$ is an allowable colouring if
$(\phi(i),\phi(j))\in E$ whenever $|i-j|=1$.  Let
$N(n)=\#\rC_{\Delta}(\an{n})$ be the cardinality on
$\rC_{\Delta}(\an{n})$.  Note that $N(n)=O(\rho(\Delta)^n)$, where
$\rho(\Delta)$ is the spectral radius of $\Delta$.  Let
$\phi,\psi\in\rC_{\Delta}(\an{n})$ be two allowable $\Delta$ colourings
of $\an{n}$.  View $(\phi,\psi)$ as a colourings of integer strip
$\tau:\an{n}\times\an{2}\subset \Z^2$, where
$\tau((i,1))=\phi(i),\tau((i,2))=\psi(i)$ for $i=1,\ldots,n$.
Let $t_{\phi,\psi}=1$ if $\tau$ is a $\Delta$-allowed colouring, i.e.
$(\phi(i),\psi(i))\in E$ for $i=1,\ldots,n$, and otherwise
$t_{\phi,\psi}=0$.  Then $T_{n,2}=[t_{\phi,\psi}]_{\phi,\psi\in
\rC_{\Delta}(\an{n})}\in \{0,1\}^{N(n)\times N(n)}$ is the transfer
matrix, which defines the allowable colourings of the infinite strip
$\an{n}\times \Z\subset \Z^2$.  An important and useful observation is
that $T_{n,2}$ is a symmetric matrix.

Let $\rho(T_{n,2})$ be the spectral radius of $T_{n,2}$.  It is well
known that the sequence $\log\rho(T_{n,2})$ is \emph{subadditive} in
$n$.  Hence the following limit exists, and it is called the
\emph{topological entropy} of $\Z^2$ with respect to $\Delta$, e.g.
\cite{Fr1,Fr2,FP}:
\begin{equation}\label{deftp2}
    h_2(\Delta):=\lim_{n\to\infty} \frac{\log\rho(T_{n,2})}{n}.
\end{equation}
Furthermore, one has the following known upper and lower
bounds, see e.g. \cite{Fr1,Fr2}
\begin{multline}\label{uplowbdsh2}
    \frac{\log\rho(T_{p+2q+1,2})-\log\rho(T_{2q+1,2})}{p}\le
    h_2(\Delta)\le \frac{\log\rho(T_{n,2})}{n}, \\ \; 1\le p,n,\; 0\le q.
\end{multline}

In many theoretical derivations, e.g. \cite{Lie}, it is useful to
consider the periodic colouring $\rC_{\Delta,\perio}(\an{n})\subset
\rC_{\Delta}(\an{n})$, where each colouring $\phi\in
\rC_{\Delta}(\an{n})$ satisfies the additional condition
$(\phi(1),\phi(n))\in E$.  So $\rC_{\Delta,\perio}(\an{n})$ is a 
$\Delta$-allowable colouring of the torus $\Z/\an{n}$.  Let
$N(n,\perio)=\#\rC_{\Delta,\perio}(\an{n})$ be the number of periodic
$\Delta$-colouring of $\rC_{\Delta,\perio}(\an{n})$.  Then
$T_{n,2,\perio}=[t_{\phi,\psi}]_{\phi,\psi\in\rC_{\Delta,\perio}(\an{n})}
\in \{0,1\}^{N(n,\perio)\times N(n,\perio)}$ is the transfer matrix
for the allowable colourings of the infinite torus $(\Z/\an{n})\times
\Z$ by $\Delta$.  Note that $T_{n,2,\perio}$ is a principal submatrix
of $T_{n,2}$, hence symmetric.  Furthermore
$\rho(T_{n,2,\perio})\le\rho(T_{n,2})$.  Also, the following
inequalities hold \cite{FP}
\begin{multline}\label{lupbdperh2}
    \frac{\log\rho(T_{p+2q,2,\perio})-\log\rho(T_{2q,2,\perio})}{p}\le
    h_2(\Delta)\le \\ \frac{\log\rho(T_{2n,2,\perio})}{2n}, 0\le q,\;1\le
    p,n,
\end{multline}
where $\rho(T_{0,2,\perio})=\rho(\Delta)$.  Note that the upper bound
in (\ref{lupbdperh2}) is an improvement of the upper bound in
(\ref{uplowbdsh2}) for an even $n$.

We now define the $1$-vertex transfer matrix
\begin{equation}\label{dim1vertm2}
    S_{n,2}=[s_{\phi,\psi}]_{\phi,\psi\in
    \rC_{\Delta}(\an{n})}\in \{0,1\}^{N(n)\times N(n)},
\end{equation}
where
\begin{multline}\label{def1vertm2}
    s_{\phi,\psi}=1 \textrm{ iff } \psi(i)=\phi(i+1) \textrm{ for }
    i=1,\ldots,n-1, \textrm{ and } \\ (\phi(1),\psi(n))\in E.
\end{multline}
$S_{n,2}$ is the transfer matrix corresponding to the $\Delta$ colouring of $\Z$
winded up on a \emph{slanted} torus, given by
$\an{n+1}\times \Z$, where the points
$(n+1,i)$ and $(1,i+1)$ are identified for each $i\in\Z$.
Note that $S_{n,2}$ is a sparse matrix, where each row and
column of $S_{n,2}$ has at most $k$ neighbours.  We show that
\begin{equation}\label{lim1versr}
    \lim_{n\to\infty} \log\rho(S_{2n+1,2})=h_2(\Delta).
\end{equation}
Thus $\log\rho(S_{n,2})$ is an approximation of $h_2(\Delta)$,
and it can be computed much faster, and with much less memory than
$\log\rho(T_{n,2})$ and $\log\rho(T_{n,2,\perio})$.
The only drawback for computing the values of
$\log\rho(S_{n,2})$, is that we do not have very sharp
inequalities of the type (\ref{uplowbdsh2}) or
(\ref{lupbdperh2}).  One of the reasons is that $S_{n,2}$
is not a symmetric matrix and can have nonreal eigenvalues.
We do have the following inequalities
\begin{multline}\label{insprad}
    \frac{\log\rho(T_{n-1,2,\perio)})}{n}\le\log\rho(S_{n,2})\le \\
    \min(\frac{\log\rho(T_{n,2})}{n},\frac{\log\rho(T_{n+1,2,\perio})}{n}),\;
    2\le n,
\end{multline}
which imply (\ref{lim1versr}).

The undirected graph $\Phi$, which can be viewed as a symmetric
digraph on two vertices
\begin{equation}\label{hardcgrph}
    \Phi=(V,E), \quad V=\{1,2\},\; E=\{(1,2),(2,1),(2,2)\},
\end{equation}
corresponds to the checkerboard constraint used in \cite{NaZ2} and hard
core lattice gas model in statistical mechanics.  (Usually $2$ is replaced by
$0$.)  We will use the $1$-vertex transfer matrix to give a heuristic
estimate of the entropy of the two dimensional hard core $h_2(\Phi)$
up to 15 digits.  Using the upper and lower bounds (\ref{uplowbdsh2})
and (\ref{lupbdperh2}) we show that the 15 digits estimates agree with
the correct values of the $h_2(\Phi)$.  Previously the value of
$h_2(\Phi)$ was known to 10 digits \cite{NaZ}.

\subsection{Overview}

We now briefly  survey the contents of the paper.  In \S2 we give the
full details of the two-dimensional isotropic symmetric case discussed
above.  In \S3 we discuss the computational results for the two-dimensional 
checkerboard constraint/hard core lattice gas.  In \S4 we
discuss the characteristic polynomial and the eigenvalues of $S_{n,2}$
for the hard core model.  In \S5 we discuss a two-dimensional
$1$-vertex transfer matrix $P_{n,2}$ for nonisotropic and nonsymmetric
colouring of $\Z^2$.  We show that $\log\rho(P_{n,2})$ is an estimate
of the two dimensional-capacity, or entropy, if the colouring of $\Z^2$ has a
friendly colour.  In \S6 we show that the results for two dimensions
apply also to the three-dimensional $1$-vertex matrix.  In the last
section we show that $1$-vertex model can also be used to estimate the
monomer-dimer model in any dimension $d$.

\section{The Two-dimensional isotropic symmetric case}
Recall that the graph $\Delta=(V,E)$, where the set of vertices $V$ is
identified with $k$ colours $\an{k}$.  We allow loops $(i,i)\in E$,
which means that two adjacent vertices of $\Z^2$ can be coloured with the
same colour $i$.  If $i\in V$ is an isolated vertex, this means that no
vertex in $\Z^2$ can be coloured in the colour $i$.  Hence we will deal
only with graphs $\Delta$ which do not have isolated vertices.

\begin{theo}\label{main1vert2d} 
    Let $\Delta=(V,E)$ be an undirected
    graph on $\#V=k\ge 2$ vertices, where self-loops are allowed, with no
    isolated vertices.  Then (\ref{insprad}) and (\ref{lim1versr}) hold.
\end{theo}
\proof  Identify the entry $s_{\phi,\psi}=1$ of $S_{n,2}$
with an allowable $\Delta$-colouring of the following $L$
shape in $\Z^2$
\begin{eqnarray}\label{deftilphc}
    &&\omega:\{(1,1),\ldots,(n,1),(1,2)\}\to
    \an{k},\\
    &&\omega((i,1))=\phi(i),\;i=1,\ldots,n,\;\omega((1,2))=\psi(n).
    \nonumber
\end{eqnarray}
More generally, consider the entries of the matrix
$S_{n,2}^m=[s_{\phi,\psi}^{(m)}]_{\phi,\psi\in\rC_{\Delta}(\an{n})}$.
Recall that
$$s_{\phi_1,\phi_{m+1}}^{(m)}=\sum_{\phi_1,\ldots,\phi_{m+1}\in
\rC_{\Delta}(\an{n+1})}\prod_{i=1}^m s_{\phi_i,\phi_{i+1}}.$$
Clearly, $\prod_{i=1}^m s_{\phi_i,\phi_{i+1}}=0,1$.  Then $\prod_{i=1}^m
s_{\phi_i,\phi_{i+1}}=1$ if and only if one has a $\Delta$-allowable
colouring $\tau\in\rC_{\Delta}(\an{n+m})$ such that
\begin{equation}\label{taucol}
    (\tau(i),\tau(j))\in E \textrm{ if } |i-j|=n.
\end{equation}
Assume that $\prod_{i=1}^m
s_{\phi_i,\phi_{i+1}}=1$.  Then we can colour the following
part of $\an{n}\times \Z$:
\begin{eqnarray*}
    (1,1),\ldots,(n,1),(1,2),\ldots,(1,q),\ldots,(l,q),\\
    q=\left\lceil \frac{n+m}{n}\right\rceil,\;l=n+m-(q-1)n,
\end{eqnarray*}
using $\tau$.  Let $\1=(1,\ldots,1)\trans$ be a vector whose
all coordinates are $1$.  The number of coordinates of $\1$
will depend on the square matrix $A=[a_{ij}]\in \R^{M\times
M}$.  Observe that $\1\trans A\1=\sum_{i,j=1}^M a_{ij}$.
Hence $\1\trans S_{n,2}^m\1$ is the number of all $\Delta$
allowable colourings $\tau\in\rC_{\Delta}(\an{n+m})$ satisfying the condition
(\ref{taucol}).  In particular, $\1\trans S_{n,2}^{(q-1)n}\1$
is the number of different colourings of the rectangle
$\an{n}\times \an{q}$ induced by $\tau\in\rC_{\Delta}(qn)$ satisfying
(\ref{taucol}).
Similarly, $\1\trans T_{n,2}^{q-1}\1$ is the number of
$\Delta$-allowable colourings of the strip $\an{n}\times{q}$.
Hence
\begin{eqnarray*}
    &&\1\trans S_{n,2}^{(q-1)n}\1\le \1\trans T_{n,2}^{(q-1)}\1
    \Rightarrow\\
    &&\log\rho(S_{n,2})=\lim_{q\to\infty} \frac{\log \1\trans
    S_{n,2}^{(q-1)n}\1}{(q-1)n}\le \\ 
    &&\lim_{q\to\infty} \frac{ \1\trans
    T_{n,2}^{(q-1)}\1}{(q-1)n}=\frac{\log\rho(T_{n,2})}{n}.
\end{eqnarray*}
See \cite[Proposition 10.1]{Fr2} for details.
This inequality establishes the first part of the upper bound in
(\ref{insprad}).  The equality (\ref{deftp2}) yields the
inequality
\begin{equation}\label{limsupin}
    \limsup_{n\to\infty} \log\rho(S_{n,2})\le h_2(\Delta).
\end{equation}
We now show the inequality $\log\rho(S_{n,2})\le
\frac{\log\rho(T_{n+1,2,\perio})}{n}$.
Let $m=(q-1)n+1$ and consider the colouring of
$\an{n}\times\an{q}\cup\{(1,q+1)\}$ induced by colouring
$\tau\in \rC_{\Delta}(\an{nq+1})$ satisfying condition
(\ref{taucol}).  Note that the number of such colourings is
$\1\trans S_{n,2}^{(q-1)n+1}\1$.
We now extend each colouring $\tau$ to the colouring
of $\an{n+1}\times\an{q}$ viewed as a colouring of the torus
$(\Z/(n+1))\times \an{q}$ as follows. The colouring of
$\an{n}\times \an{q}$ is given by $\tau$ as above.  Then the
colour of the point $(n+1,i)$ is given by $\tau(in+1)$ for
$i=1,\ldots,q$.  Hence $\1\trans S_{n,2}^{(q-1)n+1}\1 \le
\1\trans T_{n+1,2,\perio}^{q-1}\1$.  Use \cite[Proposition 10.1]{Fr2}
as above to deduce the second part of the upper bound in (\ref{insprad}).

We now show the lower bound in (\ref{insprad}).  Recall that
$\1\trans T_{n-1,2,\perio}^{q-1}\1$ is the number of colourings of the
rectangle $\omega:\an{n-1}\times\an{q}$ on the torus
$(\Z/(n-1))\times\an{q}$.  Each $\omega$ induces the following
$\tau\in\rC_{\Delta}(\an{nq})$ satisfying (\ref{taucol}):
\begin{eqnarray*}
    \tau(i+(j-1)n)=\omega((i,j)),\; \tau(j n)=\omega((1,j)),\\
    i=1,\ldots,n-1,\; j=1,\ldots,q.
\end{eqnarray*}
Hence $\1\trans T_{n-1,2,\perio}^{q-1}\1\le \1\trans
S_{n,2}^{(q-1)n}\1$.  The above arguments show the lower
inequality in (\ref{insprad}).

Combine the upper bound in (\ref{lupbdperh2}) and the lower bound
in (\ref{insprad}) to obtain
\begin{equation}\label{upbdsh2en}
    \frac{2n}{2n+1} h_2(\Delta)\le
    \frac{\log\rho(T_{2n,2,\perio})}{2n+1}\le
    \log\rho(S_{2n+1,2}).
\end{equation}
Hence $\liminf_{n\to\infty}\log\rho(S_{2n+1,2})\ge
h_2(\Delta)$.  Combine this inequality with (\ref{limsupin})
to deduce (\ref{lim1versr}).  \qed

\section{The Two dimensional run length limited channel}\label{2dchan}
We will now apply the above results to the following simple symmetric
digraph $\Phi$ on two vertices:
\begin{center}
    \psset{unit=1cm}
    \begin{pspicture}(2,.5)
	\psset{fillstyle=solid,radius=.1}
	\Cnode[fillcolor=red](0,0){red}
	\Cnode[fillcolor=blue](1,0){blue}
	\ncarc[arcangle=30]{->}{red}{blue}
	\ncarc[arcangle=30]{->}{blue}{red}
	\nccircle[angle=-90]{->}{blue}{.2}
	\Cnode[fillcolor=blue](1,0){anotherblue}
    \end{pspicture}
\end{center}
Identify the red colour with the state $1$ and the blue colour with
the state $2$, which is usually identified with the state $0$.
Then $\rC_{\Phi}(\Z^d)$ consists of all colourings of the lattice
$\Z^d$ in blue and red colours such that no two red colours are
adjacent.  In terms of codes this correspond to the checkerboard constrain used in
e.g. \cite{NaZ} and  the simplest \emph{hard core model} in
statistical mechanics.  

The adjacency matrix $A(\Phi)$ is 
$\left[\begin{array}{cc} 0 &1\\1&1\end{array}\right]$.  It is well
known that the one-dimensional entropy corresponding to
$\rC_{\Phi}(\Z)$ is the logarithm of the golden ratio
$\frac{1+\sqrt{5}}{2}$.  Hence, sometimes $\Phi$ is referred as the
golden ratio graph. 

\subsection{Computational results}
The topological entropy $h_2(\Delta)$ of the hard
core model is known within the precision of 10 digits \cite{NaZ}.  As
in the physics literature, e.g. \cite{Lie}, we will discuss the values
of $e^{h_2(\Phi)}$.  The inequalities obtained in \cite{NaZ} are
equivalent to 
\begin{eqnarray}\nonumber
    &&1.50304808247497745859985734\\
    &&<e^{h_2(\Phi)} <\label{nagzlubd}\\
    &&1.50304808257186797788004159 \nonumber
\end{eqnarray}

We have computed eigenvalues for the standard transfer matrix, the
periodic transfer matrix and finally for the 1-vertex transfer matrix.
Let us examine the results.

\begin{table*}[!ht]
    \begin{tabular}{ccc}
	$n$& $\rho(T_{n,2})$&$\rho(T_{n,2,\perio})$\\
	\hline
    2  & 2.414213562373095      & \;           \\
    3  & 3.631381260403638      &
         3.302775637731994646559610633735247 \\
    4  & 5.457705395965834      &
         5.156325174658661693523159039366916 \\
    5  & 8.203259193755024      &
         7.637519478750677316156696280583774  \\
    6  & 12.32988221531524      &
         11.55170956604814509016646221019832  \\
    7  & 18.53240737754881      &
         17.31622927332784947478739705217656 \\
    8  & 27.85509909631079      &
         26.05798609193972135567942994470689 \\
    9  & 41.8675533182809       &
         39.14578184202813825907509993927013 \\
    10  & 62.928945725187815984970517564242   &
          58.85193508152278064182392832406795\\
    11  & 94.585231204973665631062351227180   &
          88.44780432952028084071406736758034\\
    12  & 142.16615039284113705381555339180   &
          132.9477940474849517182393096863462\\
    13  & 213.68255974084561463042598863826   &
          199.8224640440179428924580367714202\\
    14  & 321.17516167688358891589859286791  &
          300.3458520273548324890314287157792\\
    15  & 482.74171089714185369639722005855   &
          451.4321236042340748832899060864991\\
    16  & 725.58400289480913382756934371728   &
          678.5256693463314967782756777455377\\
    17  & 1090.5876442258184943798276273781   &
          1019.855674771119057225300394186046\\
    18  & 1639.2056674249062545759683245949   &
          1532.892835974578082578218246454006\\
    19  & 2463.8049352057025354985027848902   &
          2304.011134761604258032847423262351\\
    20  & 3703.2172834541914402106141013810   &
          3463.039870272410387224398919252810\\
    21  & 5566.1136368853314325759144845043   &
          5205.115189611387576190086821772236\\
    22  & 8366.1364287602952611690001026376   &
          7823.538578192028601159488379441279\\
    23  & 12574.705316975186002047579953615   &
          11759.15453623673826922032891077820\\
    24  & 18900.386714371918513370098148691   &
          17674.57476300121240825274824636313\\
    25  & 28408.190009078957569653791590282   &
          26565.73564566649640784345844846860\\
    26  & 42698.875519741019043924030420844   &
          39929.57806437579207220381471608921\\
    27  & 64178.462973799644995496644648926   &
          60016.07571363818523121568140477488\\
    28  & 96463.315708983666788807112233379   &
          90207.04754031656425802350521860206\\
    29  & \;& 135585.5298162229950461965077280634\\
    30  & \;& 203791.5706122926277969020818385026\\
    31  & \;& 306308.5294259292544744183247215849\\
    32  &\; & 460396.4478048022575136612396701088\\
    33  &\; & 691997.9980476929053415088280918978\\
    34  &\; & 1040106.264045025628338539352181110\\
    35  &\; & 1563329.725741562336590052426628794\\
    36  &\; & 2349759.746553886953259135919605940\\
  \end{tabular}
  \caption{Spectral radii for standard and periodic transfer
  matrices}\label{nresh2}
\end{table*}

Table \ref{nresh2} gives the computed values of $\rho(T_{n,2}),
n=2,\ldots,28$ and $\rho(T_{n,2,\perio}), n=3,\ldots,36$.  We observed
the following facts on the two computed sequences.  First the sequence
$\frac{\rho(T_{n,2})}{\rho(T_{n-1,2})}$ is an increasing sequence for
$n\geq 10$.  Note that in view of the lower bound given in
(\ref{uplowbdsh2}) we know that
$\frac{\rho(T_{2m,2})}{\rho(T_{2m-1,2})}$ is a lower bound for
$e^{h_2(\Phi)}$.  Hence
\begin{equation}\label{exactlbh2Phi}
    1.50304808247533226432204921 \le e^{h_2(\Phi)}.
\end{equation}

The sequence $\rho(T_{2m,2,\perio})^{\frac{1}{2m}}$ is decreasing for
$m=2,\ldots,18$.  In view of the upper bound in (\ref{lupbdperh2}) it
follows that
\begin{equation}\label{exactubh2Phi}
    e^{h_2(\Phi)}\le 1.50304808247533992728837255.
\end{equation}
The above two inequalities give the correct $15$ digits of the value
of $e^{h_2(\Phi)}$.  Note that the lower bound in (\ref{nagzlubd}) is
closer to the correct value of $e^{h_2(\Phi)}$ than the upper bound.
It is plausible to assume that the same observation applies to the
inequalities (\ref{exactlbh2Phi}-\ref{exactubh2Phi}).

The sequence $\rho(T_{2m+1,2,\perio})^{\frac{1}{2m+1}}$ increases for
$m=1,\ldots,17$.  If this sequence always increases, in view of
(\ref{uplowbdsh2}), it would follows that
$\rho(T_{2m+1,2,\perio})^{\frac{1}{2m+1}}$ is a lower bound for
$e^{h_2(\Phi)}$.  The lower bound given by (\ref{exactlbh2Phi}) is
bigger than $\rho(T_{35,2,\perio})^{\frac{1}{35}}$.

\begin{table}[!ht]
    \begin{tabular}{cc}
	$n$& $\rho(S_{n,2})$\\
	\hline
	Even $n$  & \\
	26 & 1.5030480824559338746449982720899 \\
	28 & 1.5030480824713491171046098760579 \\
	30 & 1.5030480824745080695008293589330 \\
	32 & 1.5030480824751605743865692042299 \\
	34 & 1.5030480824752962878823092158144 \\
	36 & 1.5030480824753246862738777999703 \\
	38 & 1.5030480824753306606437859142329 \\
	40 & 1.5030480824753319235292607404167 \\
	Odd $n$ & \\
	39 & 1.5030480824753330032275278142102 \\
	37 & 1.5030480824753357484850986619224 \\
	35 & 1.5030480824753487657242129983806 \\
	33 & 1.5030480824754108025759894900493 \\
	31 & 1.5030480824757081424841278582465 \\
	29 & 1.5030480824771425174857112752302 \\
	27 & 1.5030480824841133358901685021830 \\
	25 & 1.5030480825182810708944214989118 \\
    \end{tabular}
    \caption{Spectral radii for $1$-vertex transfer matrices}\label{1vernresh2}
\end{table}

Table \ref{1vernresh2} gives the values of $\rho(S_{n,2})$ for
$n=25,\ldots,40$.  We found that the sequence $\rho(S_{2l,2}),
l=1,\ldots,20$ increases while the sequence $\rho(S_{2l-1,2})$
decreases for $l=2,\ldots,20$.  Assuming this behaviour for all values
$l\in\N$ we deduce that
\begin{equation}\label{decincras}
    \rho(S_{2l,2}) <
    e^{h_2(\Phi)}<\rho(S_{2l-1,2}), \textrm{ for all } l\in\N.
\end{equation}
The above \emph{ansatz} for $l=20$ gives the value of
$e^{h_2(\Phi)}$ with a precision of $15$ digits, which coincides with
the exact values given by (\ref{exactlbh2Phi}-\ref{exactubh2Phi}).
More precisely the heuristic upper bound $\rho(S_{39,2})$ is slightly
better than (\ref{exactubh2Phi}), and the heuristic lower bound given
by $\rho(S_{40,2})$ is slightly worse than (\ref{exactlbh2Phi}).

\subsection{Computational costs}
We now compare the computational resources needed in order to compute
the eigenvalues for the 1-vertex transfer matrix and the standard
transfer matrix for cycles.  The computations were done in 31 digits
precision, both were done on the same machine, and used Perron
iteration for computing the maximum eigenvalue. The standard and
periodic transfer matrices were compressed using the method of \cite{LM}
thereby reducing the amount of RAM needed considerably.
\begin{table}[ht]
    \begin{tabular}{lll}
	Matrix & Time &RAM \\
	\hline
	$N=40$ 1-vertex  & 48 CPU-hours  & 0.5GB matrix, two 4GB vectors \\
	$N=36$ periodic  & 988 CPU-hours & 203GB matrix, two 20MB vectors \\
    \end{tabular}
    \caption{Computational costs}\label{cost}
\end{table}
For the $N=40$ 1-vertex transfer matrix the computation took 48
CPU-hours, used 0.5GB RAM for the matrix and also needed two vectors
or the Perron iteration, each using 4GB RAM. For the $C_{36}$ standard transfer 
matrix the computation took 988 CPU-hours, the automorphisms compressed transfer
matrix used 203GB RAM and also needed two vectors, each using 20MB
RAM. 

We see that in order to reach a high prescision estimate of the asymptotic
capacity the 1-vertex transfer matrix used less memory and less CPU
time.  The resources needed for the largest 1-vertex transfer matrix
used here is today available on many larger workstations, while those
needed for the largest periodic transfer matrix still requires a
larger cluster.  This means that the 1-vertex transfer matrix can first
be used on a smaller inexpensive machine in order to find interesting
examples of codes so that the more costly large computer will only be
used when an interesting candidate has been found.

\section{The General Two dimensional case}
Let $\Gamma=(\an{k},E)$ be a directed graph, abbreviated as digraph,
on the set of vertices $V=\an{k}$, where $E\subset
\an{k}\times\an{k}$.  Let $X\subset \Z$.  Then $\phi:X\to \an{k}$ is
viewed as a $k$ colouring of $X$.  $\phi$ is $\Gamma$-allowable colouring
of $X$ if $(\phi(i),\phi(i+1))\in E$ whenever $i,i+1\in X$.  Denote by
$\rC_{\Gamma}(X)$ the set of allowable $\Gamma$-colourings of $X$.  Note
that a colour $j\in\an{k}$ appears in $\phi\in \rC_{\Gamma}(\Z)$ if and
only if the vertex $j$ of $\Gamma$ is contained in strongly connected
component of $\Gamma$.  Hence we assume that $G$ is a disjoint union
of strongly connected graphs.  Recall that
$\log\#\rC_{\Gamma}(\an{n}), n=1,2,\ldots,$ is a subadditive sequence.
Hence the $\Z$-entropy induced by $\Gamma$ is given by
\begin{equation}\label{ent1SOFT}
    h_1(\Gamma)=\rho(\Gamma)=\lim_{n\to\infty}
    \frac{\log\#\rC_{\Gamma}(\an{n})}{n},
\end{equation}
where $\rho(\Gamma)$ is the spectral radius of $\Gamma$,
i.e. the adjacency matrix corresponding to $\Gamma$.
See for example \cite{Fr1,Fr2}.

Let $\Gamma_1=(\an{k},E_1),\Gamma_2=(\an{k},E_2),
E_1,E_2\subseteq\an{k}\times\an{k}$ be two directed graphs on the set of
vertices $V=\an{k}$.  We assume that each $\Gamma_i$ is a disjoint
union of strongly connected graphs.  Let $X\subset \Z^2$.  For each
$i\in\Z$ denote
$$X_{1,i}=X\cap \Z\times\{i\}, \quad X_{2,i}=\{i\}\times \Z.$$

View $\phi:X\to \an{k}$ as a $k$-colouring of $X$.  Then $\phi$ is an
allowable $\Gam=(\Gamma_1,\Gamma_2)$-colouring of $X$ if $\phi|
X_{1,i},\phi|X_{2,i}$ are $\Gamma_1,\Gamma_2$-allowable colourings
respectively, for each $i\in\Z$.  Let $\rC_{\Gam}(X)$ be all $\Gam$-allowable 
colourings of $X$.  It is known that the sequence
$\log\#\rC_{\Gam}(\an{n_1}\times\an{n_2})$ is a subadditive sequence
in one variable, where the other variable is fixed.  Hence the
$\Z^2$-entropy induced by $\Gam$ is given by, e.g. \cite{Fr1,Fr2},
\begin{multline}\label{ent2SOFT}
    h_2(\Gam)=\lim_{n_1,n_2\to\infty}
    \frac{\log\#\rC_{\Gam}(\an{n_1}\times\an{n_2})}{n_1 n_2}\le\\
    \frac{\log\#\rC_{\Gam}(\an{n_1}\times\an{n_2})}{n_1 n_2}
\end{multline}
for all $n_1,n_2\in\N$.

Denote $\{i'\}=\{1,2\}\backslash\{i\}$.
Let $R_{n,i}=[r_{\phi,\psi,i}]_{\phi,\psi\in
\rC_{\Gamma_{i'}}(\an{n})}$ be the following $(0,1)$ transfer matrices
for $i=1,2$.
$r_{\phi,\psi,i}=1$ if the colouring of $\an{n}\times\an{2}$,
such that the colouring of $\{(1,1),\ldots,(n,1)\}$ and
$\{(1,2),\ldots,(n,2)\}$ is given by $\phi$ and $\psi$
respectively, is $(\Gamma_{i'},\Gamma_{i})$-allowable.
Hence
\begin{eqnarray}\nonumber
    &&\log\rho(R_{n,1})=\lim_{n_1\to\infty}
    \frac{\log\#\rC_{\Gam}(\an{n_1}\times\an{n})}{n_1}, \\
    &&\log\rho(R_{n,2})=\lim_{n_2\to\infty}
    \frac{\log\#\rC_{\Gam}(\an{n}\times\an{n_2})}{n_2},\nonumber\\
    &&\label{upbdstrmat}
    h_2(\Gam)=\lim_{n\to\infty} \frac{\log\rho(R_{n,i})}{n}\le
    \frac{\log\rho(R_{n,i})}{n}, \\
    && \quad i=1,2, \textrm{ for all }
    n\in\N.
\end{eqnarray}
Note that for general $\Gamma_1,\Gamma_2$ we do not have lower
bounds that converge to $h_2(\Gam)$.  In fact, there are
examples that the computation of $h(\Gam)$ is undecidable,
e.g. \cite{Fr1,Fr2}.  One has good lower bounds, similar to
(\ref{uplowbdsh2}), if $\Gamma_i$ is undirected for $i=1$ or
$i=2$.  Note that the isotropic case discussed in \S2 is given by
the condition that $\Gamma_1=\Gamma_2$ is an undirected graph
$\Delta$.  In that case $T_{n,2}=R_{n,1}=R_{n,2}$.

We now discuss the $1$-vertex transfer matrix for a given $\Gam$.
In this case we also have two transfer matrices $P_{n,1},P_{n,2}$
which are $(0,1)$ matrices.  For simplicity of the exposition we
discuss only $P_{n,2}$.  Let
$P_{n,2}=[p_{\phi,\psi,2}]_{\phi,\psi\in\rC_{\Gamma_1}(\an{n})}$.
Then $p_{\phi,\psi,2}=1$ if and only if the conditions given in
(\ref{def1vertm2}) hold, where $E=E_2$.  The proof of Theorem
\ref{main1vert2d} yields the inequality
\begin{equation}\label{1vertniub}
    \log\rho(P_{n,2})\le \frac{\log\rho(R_{n,2})}{n} \textrm{ for
    } n\in\N.
\end{equation}
Hence
$$h_2(\Gam)\ge \limsup_{n\to\infty} \log\rho(P_{n,2}).$$
It is an interesting question when
\begin{equation}\label{1verteq}
    h_2(\Gam)=\lim_{n\to\infty} \log\rho(P_{n,2}).
\end{equation}
Theorem \ref{main1vert2d} implies that the above equality holds if
$\Gamma_1=\Gamma_2$ is an undirected graph.  We now give
another condition for the equality (\ref{1verteq}).

A colour $j\in\an{k}$ is called a friendly colour, if $(j,i),(i,j)\in
E_1\cap E_2$ for all $i\in\an{k}$.  The existence of a
friendly colour means that $\Gamma_1$ and $\Gamma_2$ are
strongly connected graphs.
Note that for the hard core model the colour $2$ is friendly.
\begin{theo}\label{1vertfcthm}  Let
    $\Gamma_1=(\an{k},E_1),\Gamma_2=(\an{k},E_2)$ be two digraphs
    with a friendly colour.  Then for each $n\ge 1$
    \begin{equation}\label{1vertinfrcol}
	\frac{\log\rho(R_{n,2})}{n+1}\le \log\rho(P_{n+1,2}).
    \end{equation}
    Hence (\ref{1verteq}) holds.
\end{theo}
\proof As in the proof of Theorem \ref{main1vert2d} we deduce that
$$\#\rC_{\Gam}(\an{n}\times \an{m})=\1\trans
R_{n,2}^{m-1}\1.$$
Observe next that $\1\trans P_{n+1,2}^{(n+1)(m-1)}\1$ is the
number of all $\omega\in \rC_{\Gamma_1}(\an{(n+1)m})$ such that
$(\omega(i), \omega(n+1+i))\in E_2$ for $i=1,\ldots,
(m-1)(n+1)$.
Let $\tau\in\rC_{\Gam}(\an{n}\times \an{m})$.  Extend $\tau$
to $\hat \tau:\an{n+1}\times {m}$ by letting $\hat
\tau((n+1,i))=j$ for $i=1,\ldots,m$, where $j$ is a friendly
colour of $\Gam$.
Then each such $\hat \tau$ is induced by $\omega\in \rC_{\Gamma_1}(\an{(n+1)m})$
satisfying the above additional conditions.  Hence
$$\1\trans R_{n,2}^{m-1}\1\le \1\trans
P _{n+1,2}^{(n+1)(m-1)}\1.$$
The arguments of the proof of Theorem \ref{main1vert2d} yield
the inequality (\ref{1vertinfrcol}).
Combine this inequality with the inequality (\ref{1vertniub})
and the characterisation of $h_2(\Gam)$ in
(\ref{upbdstrmat}) to deduce (\ref{1verteq}).  \qed

\section{The Three dimensional case}
Let $\Gamma_i=(\an{k},E_i), E_i\subseteq \an{k}\times\an{k}$ be a
digraph, which is a disjoint union of strongly connected graphs, for
$i=1,2,3$.  Denote $\Gam=(\Gamma_1,\Gamma_2,\Gamma_3)$.  Let
$\e_i=(\delta_{1i},\delta_{i2},\delta_{i3})\trans, i=1,2,3$ be the
standard basis in $\Z^3$.  For $X\subset \Z^3$ the colouring $\phi:X\to
\an{k}$ is called $\Gam$-allowable if $(\phi(\j),\phi(\j+\e_i))\in
E_i$, whenever $\j,\j+\e_i\in X$, for $\j\in\Z^3$ and $i=1,2,3$.
Denote by $\rC_{\Gam}(X)$ all $\Gam$-allowable colourings of $X$.
Recall that the sequence
$\log\#\rC_{\Gam}(\an{n_1}\times\an{n_2}\times\an{n_3})$ is a
subadditive sequence in one variable, where the other variables are
fixed.  Hence the $\Z^3$-entropy induced by $\Gam$ is given by, e.g.
\cite{Fr1,Fr2},
\begin{eqnarray}\label{ent3SOFT}
    h_3(\Gam)=\lim_{n_1,n_2,n_3\to\infty}
    \frac{\log\#\rC_{\Gam}(\an{n_1}\times\an{n_2}\times\an{n_3})}{n_1 n_2 n_3}\\
    \le
    \frac{\log\#\rC_{\Gam}(\an{n_1}\times\an{n_2}\times\an{n_3})} {n_1
    n_2n_3}, \nonumber
\end{eqnarray}
for all $n_1,n_2,n_3\in\N$.
Let $\n=(n_1,n_2)\in\N^2$.  Denote
$\an{\n}=\an{n_1}\times\an{n_2}$.
As in the two dimensional case one
can define the transfer matrices $R_{\n,1},R_{\n,2},R_{\n,3}$ which
are $(0,1)$ matrices.
For simplicity of the exposition we discuss only
$$R_{\n,3}=[r_{\phi,\psi,3}]_{\phi,\psi\in
\rC_{(\Gamma_1,\Gamma_2)}(\an{\n})}.$$
$r_{\phi,\psi,3}=1$ if the colouring of $\an{\n}\times\an{2}$,
such that the colouring of $\an{\n}\times\{1\},
\an{\n}\times\{2\}$ given by $\phi$ and $\psi$ respectively,
is $\Gam$-allowable.  Hence
\begin{eqnarray}\nonumber
    &&\log\rho(R_{(n_1,n_2),3})=\lim_{n_3\to\infty}
    \frac{\log\#\rC_{\Gam}(\an{n_1}\times\an{n_2}\times n_3)}{n_3}, \\
    &&\label{upbdstrmat3}
    h_3(\Gam)=\lim_{n_1,n_2\to\infty} \frac{\log\rho(R_{(n_1,n_2),3})}{n_1 n_2}\le
    \frac{\log\rho(R_{(n_1,n_2),3})}{n_1 n_2},
\end{eqnarray}
for all $n_1,n_2\in\N$.

We now discuss the $1$-vertex transfer matrix for a given
$\Gam$.  In this case we have three transfer matrices
$P_{\n,1},P_{\n,2}, P_{\n,3}$ $(0,1)$ matrices.  For simplicity of the exposition we
discuss only $P_{\n,3}$.
Let $\hat\rC_{\Gamma_1}(\an{n_1n_2})$ be
the subset of all $\Gamma_1$-allowable colourings of
$\phi\in\rC_{\Gamma_1}(\an{n_1n_2})$ satisfying the additional
condition:
\begin{equation}\label{fcond}
    (\phi(i),\phi(n_1+i))\in E_2 \textrm{ for } i=1,\ldots, n_1(n_2-1).
\end{equation}
The condition (\ref{fcond}) yields that the colouring given by
$\phi$ on $\an{n_1n_2}$ induces a $(\Gamma_1,\Gamma_2)$-allowable colouring 
of $\an{\n}\times \{1\}$ as discussed in the previous section.

Let
$P_{\n,3}=[p_{\phi,\psi,3}]_{\phi,\psi\in\hat\rC_{\Gamma_1}(\an{n_1n_2})}$.
Then $p_{\phi,\psi,3}=1$ if and only if $\psi(i)=\phi(i+1)$
for $i=1,\ldots,n_1n_2-1$ and
\begin{equation}\label{scond}
    (\phi(1),\psi(n_1n_2))\in E_3.
\end{equation}
The condition (\ref{scond}) yields that if we colour
$\an{\n}\times{1}$ in $\phi$ and
$(1,1,2)$ in
the colour $\psi(n_1n_2)$, we obtain a  $\Gam$-allowable 
colouring of $\an{\n}\times\{1\}\cup\{(1,1,2)\}$.

Observe that $\1\trans
P_{\n,3}^{n_1n_2(m-1)}\1$ consists of all
$\omega\in\rC_{\Gamma_1}(n_1n_2 m)$ satisfying the above ''periodic"
conditions
\begin{enumerate}
    \item\label{fcond3}
    $(\omega(i),\omega(i+n_1))\in E_2,\;i=1,\ldots,n_1(n_2 m-1)$.
    \item\label{scond3}
    $(\omega(i),\omega(i+n_1n_2)\in E_3,\;i=1,\ldots,n_1n_2(m-1)$.
\end{enumerate}
Hence each such $\omega$ induces a $\Gam$-allowable colouring of
$\an{\n}\times\an{m}$.  Therefore
$$\1\trans P_{\n,3}^{n_1n_2(m-1)}\1\le \1\trans
R_{\n,3}^{m-1}\1.$$
The proof of Theorem
\ref{main1vert2d} yields the inequality
\begin{equation}\label{1vertniub3}
    \log\rho(P_{(n_1,n_2),3})\le \frac{\log\rho(R_{(n_1,n_2),3})}{n_1n_2} \textrm{ for
    } n_1,n_2\in\N.
\end{equation}
Hence
$$h_3(\Gam)\ge \limsup_{n_1,n_2\to\infty} \log\rho(P_{\n,3}).$$
It is an interesting question when
\begin{equation}\label{1verteq3}
    h_3(\Gam)=\lim_{n_1,n_2\to\infty} \log\rho(P_{(n_1,n_2),3}).
\end{equation}

A colour $j\in\an{k}$ is called a friendly colour, if $(j,i),(i,j)\in
E_1\cap E_2\cap E_3$ for all $i\in\an{k}$.
\begin{theo}\label{1vertfcthm3}  Let
    $\Gamma_1=(\an{k},E_1),\Gamma_2=(\an{k},E_2),\Gamma_3$ be three
    digraphs with a friendly colour.  Then for each $n_1,n_2\ge 1$
    \begin{equation}\label{1vertinfrcol3}
	\frac{\log\rho(R_{(n_1,n_2),3})}{(n_1+1)(n_2+1)}\le \log\rho(P_{(n+1,n_2+1),3}).
    \end{equation}
    Hence (\ref{1verteq3}) holds.
\end{theo}
\proof  Let $\phi\in\rC_{\Gam}(\an{n_1}\times\an{n_2}\times
\an{m})$.  Extend $\phi$ to the colouring $\hat
\phi:\an{n_1+1}\times\an{n_2+1}\times\an{m}$ by colouring
the additional points in a friendly colour.  Note that
$\hat\phi$ induces the colouring
$\omega\in\rC_{\Gamma_1}((n_1+1)(n_2+1)m)$ such that
\begin{eqnarray}\nonumber
    (\omega(i),\omega(i+n_1+1))\in E_2, \\
    i=1,\ldots,
    (n_1+1)(n_2+1)m-n_1-1,\\
    \label{perconn1n2}\\
    (\omega(j),\omega(j+(n_1+1)(n_2+1)))\in E_3, \\
    j=1,\ldots
    (n_1+1)(n_2+1)(m-1).\nonumber
\end{eqnarray}
Hence
$$\1\trans R_{(n_1,n_2),3}^{m-1}\1\le \1\trans
P_{(n_1+1,n_2+1),3}^{(n_1+1)(n_2+1)(m-1)}\1.$$
Therefore (\ref{1vertinfrcol3}) holds.
Use (\ref{1vertniub3}) and (\ref{upbdstrmat3}) to deduce
(\ref{1verteq3}).  \qed

We now discuss briefly the isotropic symmetric case, i.e.
$\Gamma_1=\Gamma_2=\Gamma_3=\Delta=(\an{k},E)$ is an undirected graph
with no isolated vertices.  Then the transfer matrix
$T_{(n_1,n_2),3}$ is equal to $R_{(n_1,n_2),3}$.  Denote by
$T_{(n_1,n_2),3,\perio}$ the transfer matrix corresponding to
all $(n_1,n_2)$ periodic colourings, which are $\Delta$-allowable
colourings of $(\Z/n_1)\times (\Z/n_2)$.
Note that $T_{(n_1,n_2),3,\perio}$ is a principal submatrix of
$T_{(n_1,n_2),3}$.  Recall the following improvement of the
(\ref{upbdstrmat3}) \cite{Fr1,Fr2,FP}
\begin{multline}\label{heupperbd3}
    h_3(\Gam)=\lim_{n_1,n_2\to\infty} \frac{\log\rho(T_{(2n_1,2n_2),3,\perio})}{4n_1 n_2}\le
    \\
    \frac{\log\rho(T_{(2n_1,2n_2),3,\perio})}{4n_1 n_2},
\end{multline}
for all $n_1,n_2\in\N$.

Let $S_{(n_1,n_2),3}=P_{(n_1,n_2),3}$ be the $1$-vertex
transfer matrix corresponding to $\Delta$.
A given $\Delta$-allowable colouring on $\Z\times\Z\times\an{m}$,
which is $(n_1,n_2)$ periodic on each $\Z\times\Z\times\{i\}$
for each $i\in\an{m}$, induces a colouring $\phi\in
\rC_{\Delta}(\an{n_1+1}\times\an{n_2+1}\times\an{m})$.
This colouring corresponds to the colouring
$\omega\in\rC_{\Delta}((n_1+1)(n_2+1)m)$ that satisfies
the conditions (\ref{perconn1n2}), where $E_2=E_3=E$.
Hence we have an analogue of (\ref{1vertinfrcol3})
\begin{equation}\label{1vertinfrcol3iso}
    \frac{\log\rho(T_{(n_1,n_2),3,\perio})}{(n_1+1)(n_2+1)}\le \log\rho(S_{(n_1+1,n_2+1),3}).
\end{equation}
Use (\ref{heupperbd3}), (\ref{1vertniub3}) and
(\ref{upbdstrmat3}) to deduce
\begin{equation}\label{lim1versr3}
    \lim_{n_1,n_2\to\infty} \log\rho(S_{(2n_1+1,2n_2+1),3})=h_3(\Delta).
\end{equation}

\section{The Three-dimensional Run Length Limited Channel}
The capacity of the three-dimensional channel corresponding to the
isotropic symmetric case of $\Z^{3}$ with the same digraph as in
Section \ref{2dchan} is the three-dimensional $(0,1)$ run length
limited channel.  This channel was considered in \cite{NaZ2} where the
transfer matrix methods previously used for the two-dimensional case
was extended to find the bounds
$$ 1.43644 \leq e^{h_3(\Gam)}\leq 1.44082$$

This case has also been studied in the physics literature, see e.g.
\cite{FLA}, in the form of a three-dimensional lattice gas, where
colourings are weighted according to their number of 1s.  Mathematical
results one the structure of the set of colourings have shown
\cite{G,GK} that many of the ordinary Monte Carlo algorithms used to
study this model numerically have slow mixing properties and that
there are interesting long range correlations in the positions of the
1s.

\subsection{Computational results}
We first used the standard transfer matrix method in combination with 
the compression method of \cite{LM} to compute the largest eigenvalue
for both the periodic and aperiodic graphs needed to bound the
capacity for this channel. In order to verify our program we
recomputed the eigenvalues from \cite{NaZ2} and the results agreed. 
In Table \ref{3dtab1} we show the eigenvalues not previously computed. 
\begin{table*}[!ht]
    \begin{tabular}{lllll}
	$n_{1}$ & $n_{2}$ & aperiodic            &  periodic         & aperiodic/periodic\\ 
	\hline
	3       & 12      &                      &                   & 769999.6006127802 \\
	\hline
	4       & 7       & 41543.31662520356    &                   & \\
	4       & 8       & 184654.5439467464    & 117151.9963311473 & \\
	4       & 10      &                      & 215347.9226316121 & \\
	\hline
	5       & 5       & 13427.06985344107    & 8185.111027254276 & 10331.06553679985 \\
	5       & 6       & 85738.84889761954    & 51133.96879131764 & 68793.17993810770 \\
	5       & 7       & 547489.0319884565    & 307078.2654460451 &                   \\
	5       & 8       &                      &                   & 2779905.231816480 \\
	\hline
	6       & 6       & 786528.5060953929    & 482862.3074476483 & \\
	6       & 8       &                      & 37133338.84386827 & \\
    \end{tabular}
    \caption{Largest eigenvalue for the standard transfer matrix}\label{3dtab1}
\end{table*}
Using these eigenvalues as in \cite{NaZ,FP} we find the following
bounds for the exponentiated capacity:
\begin{equation}\label{3dbound} 
    1.4365871627266 \leq e^{h_3(\Gam)}\leq 1.43781634614
\end{equation}
We have here determined one more decimal in the value of
$e^{h_3(\Gam)}$, and limited the next decimal to only two possible
values.

The largest eigenvalues for the 1-vertex transfer are given in
Table \ref{3dtab2}. As we can see the  eigenvalues give a reasonably
good approximation of the capacity. Even for a small case like (4,4)
the first three decimals agree with the bounds in \ref{3dbound}. For
the largest case the eigenvalue is in the interval given in
\ref{3dbound}, but here we do not have an error bound. At first sight 
the values given by the 1-vertex transfer matrix looks less impressive
for this case, but in fact the best bounds before the current paper
only gave  $e^{h_3(\Gam)}=1.4$. Hence even the $(4,4)$ case for the
1-vertex transfer matrix gave an estimate with improved accuracy.
\begin{table}[!ht]
    \begin{tabular}{lll}
	$n_{1}$ & $n_{2}$ &   \\ 
	\hline
	4 & 4 & 1.431707\\
	4 & 5 & 1.433880 \\
	5 & 4 & 1.433943 \\
	5 & 5 & 1.439764 \\
	6 & 5 & 1.436801 \\
    \end{tabular}
    \caption{Largest eigenvalue for the 1-vertex transfer matrix}\label{3dtab2}
\end{table}

\subsection{Computational costs}
The most striking feature for the three-dimensional case is the
 difference in computational resources needed for the two approaches.

The computation of the eigenvalues for $C_{6,8}$ used 3700 GB RAM and
ran for 10.6 hours on 512 4-core CPUs on a linux cluster, giving a
total of about 21700 CPU-hours.  The program was a Fortan 90 program
using OpenMP and MPI for communication. Before compression the matrix 
had side 1682382.

For the 1-vertex transfer matrix constraints, in programming time and 
computer access, made us settle for
a single machine implementation of the method in Mathematica which was
run on a desktop Macintosh with 2GB RAM. 
For $n_{1}=6,n_{2}=5$ the computation used 52 CPU-hours. The matrix
has side 339 000 but since it has only two non-zero positions in each
row the amount of RAM used was negligible.
\begin{table}[ht]
    \begin{tabular}{lll}
	Matrix & Time &RAM \\
	\hline
	$n_{1}=6, n_{2}=5$ 1-vertex  & 52 CPU-hours   & 5MB matrix  \\
	$C_{6}\times C_{8}$ periodic & 21700 CPU-hours& 3700GB matrix \\
    \end{tabular}
    \caption{Computational costs for three-dimensions}\label{cost3}
\end{table}

\section{The monomer-dimer model}
The classical monomer-dimer model on $\Z^d$ consists of tiling
$\Z^d$ with monomers and dimers in the direction
$\e_i=(\delta_{1i},\ldots,\delta_{di})\trans$ for
$i=1,\ldots,d$.  This tiling can be coded in $2d+1$ colours,
see \cite{Fr2,FP}.  That is, the colour $2d+1$ corresponds to a monomer,
and the colours $2i-1,2i$ correspond
a dimer in the direction $\e_i$.  So
$\Gamma_i=(\an{2d+1},E_i)$ is given by the following
conditions on $E_i$ for $i=1,\ldots,d$.
\begin{itemize}
    \item  All vertices $\an{2d+1}\backslash\{2i-1,2i\}$ form a complete directed
    graph on $2d-1$ vertices.
    \item  $(j,2i-1), (2i,j)\in E_i$ for all
    $j\in\an{2d+1}\backslash\{2i-1,2i\}$.
    \item $(2i-1,2i)\in E_i$.
\end{itemize}

Denote by $h_d$ the monomer-dimer entropy, e.g. \cite{Fr2,FP}.
Let $\n=(n_1,\ldots,n_{d-1})\in \N^{d-1}$.  Define the
$1$-vertex transfer matrix $P_{\n,d}$ as in the previous
sections for $d=2,3$.
We claim that
\begin{equation}\label{1verteqd}
    h_d=\lim_{n_1,\ldots,n_{d-1}\to\infty} \log\rho(P_{(n_1,\ldots,n_{d-1}),d}).
\end{equation}
In this case the colour $2d+1$ is not friendly.  It is shown in
\cite{FP} that to compute the $d$-dimer entropy $h_d$ it is
enough to consider the tiling of the box $\phi$ of
$\an{n_1}\times\ldots\times\an{n_d}$ with monomers and dimers
that fit inside the box.  In this case we can extend the tiling $\phi$
to the tiling $\tilde\phi$ of $\an{n_1+1}\times\ldots\times\an{n_{d-1}+1}
\times\an{n_d}$ by adding an additional layer of monomers.  The
arguments of the previous section yield the equality
(\ref{1verteqd}).

\section*{Acknowledgements}
This research was conducted using the resources of High Performance
Computing Center North (HPC2N).

\newcommand{\etalchar}[1]{$^{#1}$}
\providecommand{\bysame}{\leavevmode\hbox to3em{\hrulefill}\thinspace}
\providecommand{\MR}{\relax\ifhmode\unskip\space\fi MR }
\providecommand{\MRhref}[2]{%
  \href{http://www.ams.org/mathscinet-getitem?mr=#1}{#2}
}
\providecommand{\href}[2]{#2}


\begin{thebibliography}{BCL{\etalchar{+}}06}

\bibitem[BCL{\etalchar{+}}06]{BCL}
Christian Borgs, Jennifer Chayes, L{\'a}szl{\'o} Lov{\'a}sz, Vera~T. S{\'o}s,
  and Katalin Vesztergombi, \emph{Counting graph homomorphisms}, Topics in
  discrete mathematics, Algorithms Combin., vol.~26, Springer, Berlin, 2006,
  pp.~315--371. \MR{MR2249277 (2007f:05087)}

\bibitem[BET80]{BET}
R.~J. Baxter, I.~G. Enting, and S.~K. Tsang, \emph{Hard-square lattice gas}, J.
  Statist. Phys. \textbf{22} (1980), no.~4, 465--489. \MR{MR574008 (81f:82037)}

\bibitem[CW98]{CW}
Neil~J. Calkin and Herbert~S. Wilf, \emph{The number of independent sets in a
  grid graph}, SIAM J. Discrete Math. \textbf{11} (1998), no.~1, 54--60
  (electronic). \MR{MR1612853 (99e:05010)}

\bibitem[FLS07]{FLS}
Michael Freedman, L{\'a}szl{\'o} Lov{\'a}sz, and Alexander Schrijver,
  \emph{Reflection positivity, rank connectivity, and homomorphism of graphs},
  J. Amer. Math. Soc. \textbf{20} (2007), no.~1, 37--51 (electronic).
  \MR{MR2257396 (2007h:05153)}

\bibitem[FP05]{FP}
Shmuel Friedland and Uri~N. Peled, \emph{Theory of computation of
  multidimensional entropy with an application to the monomer-dimer problem},
  Adv. in Appl. Math. \textbf{34} (2005), no.~3, 486--522. \MR{MR2123547
  (2005m:82020)}

\bibitem[Fri97]{Fr1}
Shmuel Friedland, \emph{On the entropy of $\mathbf{Z}^d$ subshifts of finite
  type}, Linear Algebra Appl. \textbf{252} (1997), 199--220. \MR{MR1428636
  (98b:54052)}

\bibitem[Fri03]{Fr2}
\bysame, \emph{Multi-dimensional capacity, pressure and {H}ausdorff dimension},
  Mathematical systems theory in biology, communications, computation, and
  finance ({N}otre {D}ame, {IN}, 2002), IMA Vol. Math. Appl., vol. 134,
  Springer, New York, 2003, pp.~183--222. \MR{MR2043239 (2004m:37027)}

\bibitem[Gal08]{G}
David Galvin, \emph{Sampling independent sets in the discrete torus}, Random
  Structures Algorithms \textbf{33} (2008), no.~3, 356--376. \MR{MR2446486}

\bibitem[GK04]{GK}
David Galvin and Jeff Kahn, \emph{On phase transition in the hard-core model on
  {$\mathbf{Z}^d$}}, Combin. Probab. Comput. \textbf{13} (2004), no.~2,
  137--164. \MR{MR2047233 (2005c:82034)}

\bibitem[KZ99]{KZ}
Akiko Kato and Kenneth Zeger, \emph{On the capacity of two-dimensional
  run-length constrained channels}, IEEE Trans. Inform. Theory \textbf{45}
  (1999), no.~5, 1527--1540. \MR{MR1699075 (2000c:94012)}

\bibitem[{Lie}67]{Lie}
E.~H. {Lieb}, \emph{{Residual Entropy of Square Ice}}, Physical Review
  \textbf{162} (1967), 162--172.

\bibitem[LM08]{LM}
Per~H{\aa}kan Lundow and Klas Markstr{\"o}m, \emph{Exact and approximate
  compression of transfer matrices for graph homomorphisms}, LMS J. Comput.
  Math. \textbf{11} (2008), 1--14. \MR{MR2379936 (2009e:05192)}

\bibitem[MLA07]{FLA}
H.~C. {Marques Fernandes}, Y.~{Levin}, and J.~J. {Arenzon}, \emph{{Equation of
  state for hard-square lattice gases}}, Physical Review E \textbf{75} (2007),
  no.~5, 052101--+.

\bibitem[NZ99]{NaZ}
Zsigmond Nagy and Kenneth Zeger, \emph{Capacity bounds for the 3-dimensional
  {$(0,1)$} runlength limited channel}, Applied algebra, algebraic algorithms
  and error-correcting codes ({H}onolulu, {HI}, 1999), Lecture Notes in Comput.
  Sci., vol. 1719, Springer, Berlin, 1999, pp.~245--251. \MR{MR1846500
  (2002i:94023)}

\bibitem[NZ03]{NaZ2}
\bysame, \emph{Asymptotic capacity of two-dimensional channels with
  checkerboard constraints}, IEEE Trans. Inform. Theory \textbf{49} (2003),
  no.~9, 2115--2125. \MR{MR2004767 (2005g:94061)}

\end{thebibliography}
\end{document}